\documentclass[twocolumn,aps,pre,epsf,nofootinbib,superscriptaddress]{revtex4-2}

\usepackage{amsmath}
\usepackage{amssymb}
\usepackage{epsfig}
\usepackage[T1]{fontenc}
\usepackage[utf8]{inputenc}
\usepackage{color}
\usepackage[normalem]{ulem}
\usepackage{tikz}
\usetikzlibrary{positioning}
\usepackage{tabularx}
\usepackage{enumitem}
\usepackage{graphicx}
\usepackage{subcaption}
\usepackage{ragged2e}
\DeclareCaptionJustification{justified}{\justifying}
\usepackage{dcolumn}
\usepackage{bm}
\usepackage{comment}
\usepackage{siunitx}
\usepackage{caption}
\usepackage[title]{appendix}
\usepackage{booktabs}
\usepackage{float}
\usepackage{color}
\usepackage{booktabs}
\usepackage{multirow}
\usepackage{lipsum}
\usepackage{tabulary,xcolor}
\usepackage{amsfonts}
\usepackage[T1]{fontenc}
\usepackage{verbatim} 
\usepackage{url,morefloats,floatflt,cancel}
\usepackage{graphicx}

\begin{document}

\title{Dynamical critical behavior of the two-dimensional three-state Potts model}

\author{Erol Vatansever}
\email{erol.vatansever@deu.edu.tr}
\affiliation{Department of Physics, Dokuz Eyl\"{u}l University, TR-35160, Izmir, Turkey}

\author{Gerard T. Barkema}
\email{G.T.Barkema@uu.nl}
\affiliation{Department of Information and Computing Sciences, Utrecht University, Princetonplein 5, 3584 CC Utrecht, the Netherlands}

\author{Nikolaos G. Fytas}
\email{nikolaos.fytas@essex.ac.uk}

\affiliation{School of Mathematics, Statistics and Actuarial Science, University of Essex, Colchester CO4 3SQ, United Kingdom}

\date{\today}

\begin{abstract}
We investigate the dynamical critical behavior of the two-dimensional three-state Potts model with single spin-flip dynamics in equilibrium. We focus on the mean-squared deviation of the magnetization $M$ (MSD$_{M}$) as a function of time, as well as on the autocorrelation function of $M$. Our simulations reveal the existence of two crossover behaviors at times $\tau_1 \sim L^{z_1}$ and $\tau_2 \sim L^{z_2}$, separating three dynamical regimes. MSD$_{M}$ appears to shift from ordinary diffusion in the first regime, to anomalous diffusion in the second, and finally to be constant in the third regime. The magnetization autocorrelation function on the other hand is found to fluctuate between exponential decay, stretched-exponential decay, and then again exponential decay along these three regimes. This behavior is in agreement with the one reported recently for the two-dimensional Ising ferromagnet [Phys. Rev. E {\bf 108}, 034118 (2023)], indicating that the existence of two dynamic critical exponents is not a peculiarity of the Ising model itself. A comparison of both MSD$_{M}$ and the magnetization's autocorrelation function suggests that within our numerical accuracy the exponents $z_1$ and $z_2$ are shared between the Ising and three-state Potts models at least for the particular case of single spin-flip dynamics studied here, even though their equilibrium universality classes are clearly distinct. Continuity of MSD$_{M}$ requires that $\alpha (z_2 - z_1) = \gamma/\nu - z_1$, in which $\alpha$ is the anomalous exponent in the intermediate regime. Since the ratio $\gamma/\nu$ is not shared between the two  models, it follows that $\alpha$ is not shared either, an aspect well verified in our simulations. Finally, we also discuss the relevance of our main findings using another useful observable, namely the line magnetization $M_{l}$. 
\end{abstract}

\maketitle

\section{Introduction}
\label{sec:intro}

Universality, erstwhile phenomenologically established, has been a leading principle of critical phenomena~\cite{fisher74,fisher98}. It is the property that models or systems can have the exact same set of critical exponents, describing their behaviour near a critical point of a second-order phase transition, even though their microscopic properties are very different. The explanation of universality, in terms of diverse Hamiltonian flows to a single fixed point, has been one of the crowning achievements of renormalization-group theory~\cite{wilson71}.
There are several spin models in this context which have been used for the identification and classification of universality classes~\cite{landau_book,barkema_book,amit_book}. Two of the most common are the Ising model~\cite{Ising25}, the simplest fruit-fly model of statistical physics, and the $q$-state Potts model~\cite{potts} which showcases a rich critical behavior depending on the number of spin states $q$~\cite{wu82}. Of course, for $q=2$ the Ising case is recovered. 

Although the existence and quantitative description of universality classes in most pure spin systems with simple interactions is well-established for equilibrium properties, the situation regarding dynamical properties is much more involved~\cite{hohenberg77,folk06,hasenbusch07,zhong20}. Fortunately however, the concepts of critical phenomena can be broadly speaking extended to dynamical processes~\cite{hohenberg77} and this has already pushed forward our understanding of the field. 

In a numerical study, the dynamics of a lattice model can be probed in various settings. One might consider the autocorrelation time $\tau$ of a system in equilibrium or various off-equilibrium situations. Roughly speaking, the autocorrelation time $\tau$ is the time needed to generate a statistically independent configuration in a stochastic process at equilibrium. In the neighborhood of a critical point the autocorrelation time increases with increasing correlation length $\xi$, a phenomenon called critical slowing down. The increase is governed by a power law of the form $\tau \sim \xi^{z}$~\cite{nightingale96,hasenbusch20}, where $z$ is the dynamic critical exponent, an exponent which is unrelated to the the static exponents. Note that in a finite system, $\xi$ is bounded by the linear system size $L$, so that $\tau \sim L^{z}$ at the incipient critical point. The exponent $z$ is the main critical entity that defines the corresponding dynamical universality classes, even in very complicated models, such as spin glasses~\cite{janus}.

Recently, a new window of opportunities in exploring further the universality aspects of dynamical phenomena has been opened~\cite{liu23}. Via extensive simulations of the square-lattice Ising ferromagnet it was shown that, in contrast to the standard belief of a single dynamic exponent (denoted as $z_2$ in our framework), there is another dynamic critical exponent $z_1$ which also appears to be of great practical relevance, since for obtaining statistically uncorrelated samples the proper sampling frequency should be set by the newly introduced exponent. We should underline here that earlier work on non-equilibrium dynamics has also suggested the presence of a new exponent $\theta$~\cite{janssen89} akin to the newly introduced exponent $z_1$ in Ref.~\cite{liu23}. Our previous work communicated that the dynamical critical behavior of the Ising model with Glauber dynamics is much richer than recorded so far, thus calling for further examination of dynamical universality classes, even in simple models.

In this context, the three-state Potts model is a natural upcoming candidate that has been anyway extensively used in the literature of critical phenomena and that lies in a different equilibrium universality class as compared to that of the Ising ferromagnet. We note the static critical exponent values $\nu = 5/6$, $\beta/\nu = 2/15$, and $\gamma/\nu = 26/15$ for the three-state Potts universality class~\cite{wu82} and the values $\nu = 1$, $\beta/\nu = 1/8$, and $\gamma/\nu = 7/4$ for the Ising one. From a dynamics perspective, although the dynamic critical exponent $z$ of the Ising model at two dimensions is known without doubt by the seminal work of Nightingale and Bl\"ote to be $z = 2.1665(12)$~\cite{nightingale96}, the same is not true for the three-state Potts model on the square lattice. In particular, for the latter model there is currently no consensus on the value of $z$, and the estimates suggested in the literature, using a variety of methods from Monte Carlo simulations to dynamic renormalization group and short-time scaling, span the wide range $z = 2.17(4) - 2.7(4)$~\cite{tobochnik81,arcangelis86,tang87,schulke95,schulke96,murase08}.

In the current paper we attempt to enrich our understanding  of dynamical critical phenomena by presenting a substantiative analysis on the dynamical behavior of the two-dimensional three-state Potts model. Following the prescription of Ref.~\cite{liu23}, we focus on the mean-squared deviation of the magnetization $M$, MSD$_{M}$, as a function of time, as well as on the autocorrelation function of $M$. Our simulations manifest the existence of two crossover behaviors at times $\tau_1 \sim L^{z_1}$ and $\tau_2 \sim L^{z_2}$, separating three dynamical regimes. This behavior evinces the existence of two dynamic critical exponents $z_1$ and $z_2$, as in the case of the simple Ising ferromagnet. Our main finding is that the dynamical critical behavior appears to be independent of the equilibrium universality class for the present models under study and the particular choice of local single spin-flip dynamics and that the dynamic exponents $z_1$ and $z_2$ are not determined by their corresponding static ones, such as $\gamma$ and $\nu$. Although we can not make any more general claims regarding transversal universality across models belonging to different equilibrium universality classes we believe that our analysis brings to light an interesting feature in dynamical critical phenomena that has not been previously reported. Finally, we also provide a complementary analysis in the same framework based on the line magnetization $M_{l}$ of the system, from which additional important conclusions may be drawn.

The rest of the paper is organized as follows: In Sec.~\ref{sec:model} we define the Potts model and give an outline of the numerical protocol. Subsequently in Sec.~\ref{sec:results} and after introducing the basic observables, we present our numerical data for the bulk and line magnetizations, providing insightful comparisons with the already obtained results for the Ising ferromagnet. This contribution closes with a summary and some concluding remarks in Sec.~\ref{sec:summary}.

\section{Model and Simulation Details}
\label{sec:model}

Two versions of the Potts model exist, often referred to as the {\it vector} and {\it standard} Potts model. For $q = 2$ and $3$ the two models are identical.
We consider here the nearest-neighbor, zero-field, three-state {\it vector} Potts model (as it has a more natural magnetization definition), described by the Hamiltonian~\cite{wu82}
\begin{equation}\label{eq:potts_hamiltonian}
\mathcal{H} = -\frac{2J}{3} \sum_{\langle ij \rangle} \vec{\sigma}_i\cdot \vec{\sigma}_j.
\end{equation}
In the above equation $J > 0$ indicates ferromagnetic interactions, $\left\langle \ldots \right\rangle$ refers to summation over nearest neighbors only on the square lattice, and $\vec{\sigma}_i$ defines the spin vector on lattice site $i$ which can take three possible options: $(-\frac{1}{2}, \frac{1}{2} \sqrt{3})$, $(-\frac{1}{2}, -\frac{1}{2} \sqrt{3})$, or $(1,0)$. Note that $\vec{\sigma}_i\cdot \vec{\sigma}_j = 1$ for identical spin vectors, and $-\frac{1}{2}$ otherwise. This indicates that the Hamiltonian~\eqref{eq:potts_hamiltonian} is mathematically equivalent to its more conventional version, where the spins take an integer value ($0$, $1$, or $2$) and the dot-product is replaced by the Kronecker delta-function (apart from a constant offset in the energy). Many equilibrium properties of this model are known, such as the exact location of the critical temperature, i.e., $T_{\rm c} = 1/\ln(1+\sqrt{3})$~\cite{potts} but also its critical exponents~\cite{wu82}, as also outlined in the previous Section.

The Potts model is a generalization of the Ising model and in principle its dynamics can also be a straightforward extension of Glauber dynamics~\cite{martinelli99,randall00,coulon04}. 
However in the present work we chose for simplicity to implement the heat-bath algorithm, where an elementary move is a proposed change of a single spin at a random location, which is then accepted or rejected according to the heat-bath acceptance ratio~\cite{barkema_book}. One unit of time then consists of $N = L^2$ elementary moves, where $L$ denotes the linear dimension of the lattice. Other commonly used dynamical algorithms in the extensive literature are the spin-exchange (Kawasaki) dynamics~\cite{grandi96,smedt03,godreche04}, as well as numerous types of cluster algorithms~\cite{coddington92,rieger99,bloete02}. Yet, these are outside the scope of the current work. 

Our numerical simulations of the three-state Potts model were performed at the exact critical temperature~\cite{potts} using single spin-flip dynamics of heat-bath type and systems with linear sizes within the range $L = \{16 - 192\}$, employing periodic boundary conditions. As a reference, the simulation time needed for a single realization of a system with linear size $L = 96$ on a node of \textit{Dual Intel Xeon E5-2690 V4} processor was $\sim 80$ minutes. To ensure a sufficiently good statistical accuracy in our numerical data, an extensive averaging over $10^{4} - 10^{5}$ independent realizations was performed for all sizes studied. 

\section{Results and Analysis}
\label{sec:results}

\subsection{Bulk magnetization}
\label{sec:bulk}

\begin{figure}
\includegraphics[width=8.0cm]{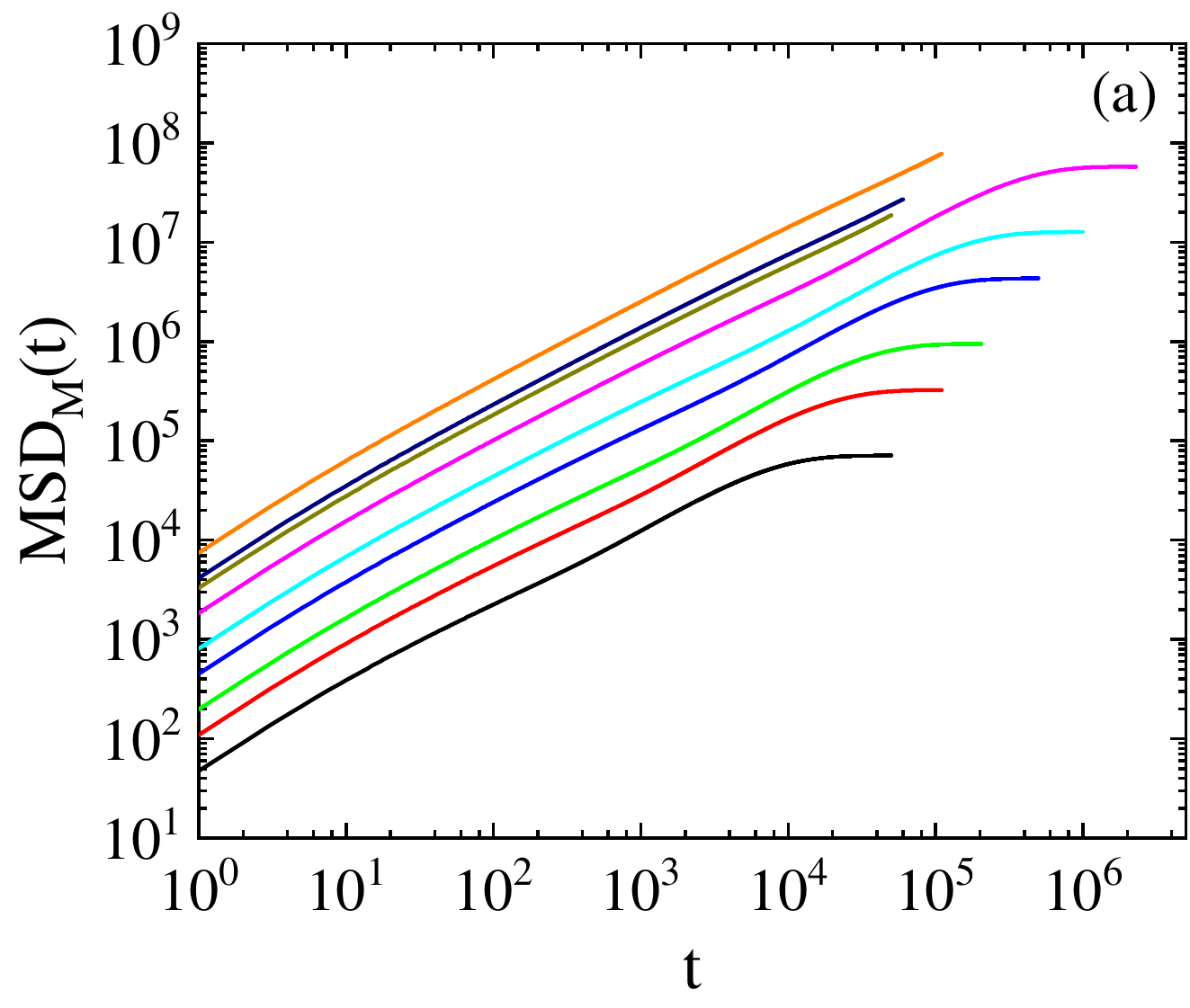}
\includegraphics[width=8.0cm]{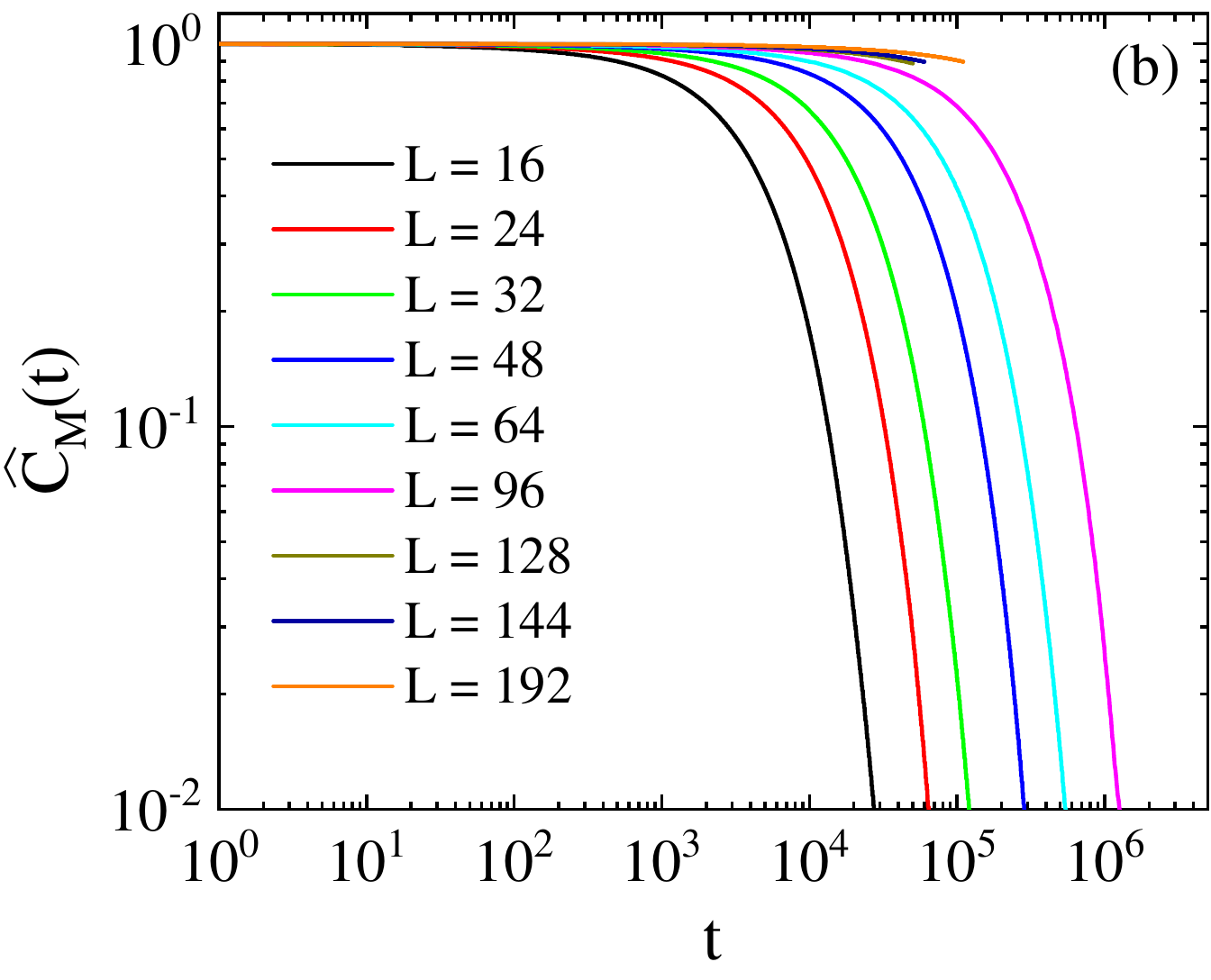}
\caption{\label{Fig1}
(a) Mean-square displacement of the magnetization MSD$_M(t)$ and (b) normalized autocorrelation function $\hat{C}_M(t)$ vs. time ($t$) for the same set of system sizes given in panel (b).}
\end{figure}

We follow in this Section the analysis performed for the Ising ferromagnet in Ref.~\cite{liu23}. The two key observables that allow us to elaborate on some new aspects of the dynamical behavior of the Potts model are based on the order parameter (bulk magnetization) of the system, $\vec{M} = \sum_i \vec{\sigma}_i$, as it fluctuates in its equilibrium state. The first corresponds to the mean-squared deviation of the magnetization
\begin{equation}
\text{MSD}_M(t)=\left \langle \left| \Delta \vec{M}(t) \right|^2 \right \rangle  = \left \langle \left| \vec{M}(t)-\vec{M}(0) \right|^2 \right \rangle,
	\label{eq:msd}
\end{equation}
and the second to the magnetization's autocorrelation function, defined as usual via
\begin{equation}
	C_M(t)=\langle \vec{M}(t) \cdot \vec{M}(0) \rangle.
	\label{eq:autocor}
\end{equation}

Figure~\ref{Fig1} summarizes the raw numerical data obtained for the two-dimensional three-state Potts model. In particular, Fig.~\ref{Fig1}(a) depicts the MSD$_M(t)$, whereas Fig.~\ref{Fig1}(b) the normalized autocorrelation $\hat{C}_M(t)=\langle \vec{M}(t) \cdot \vec{M}(0) \rangle/ \langle \left| \vec{M}(0) \right| ^2 \rangle$, both as a function of time. Three well-defined regimes are detected, separated by two crossover correlation times, which we define hereafter as $\tau_1$ and $\tau_2$, respectively.

At short times $t$, the dynamics contains $L^2 t$ proposed spin flips at spatially separated locations, out of which the numerically computed fraction $f_p\approx 0.33$ is accepted. The dynamics thus involve $f_p L^2 t$ uncorrelated changes of $\left| \Delta M \right|=\pm \sqrt {\frac{3}{2}}$. That being so, MSD$_M$ in the short-time regime is described by
\begin{equation}
	\text{MSD}_M= \frac{3}{2} f_p L^2 t \quad (t \ll \tau_1).
\end{equation}
At these short times, the magnetization does not have enough time to change significantly, remaining close to its $t = 0$ value. Additionally, the expectation of the squared magnetization is related to the magnetic susceptibility~\cite{barkema_book}
\begin{equation}
	\chi=\frac{\beta}{L^2} \langle M^2 \rangle.
\end{equation}
Thus, in the short-time regime (using the equilibrium property $\chi \sim L^{\gamma/\nu}$), we have that 
\begin{equation}
	C_M(t) \approx k_b T L^2 \chi \sim L^{2+\gamma/\nu} \quad (t \ll \tau_1).
\end{equation}

Conversely, at very long times now, the two values of the magnetization are uncorrelated so that $\langle \vec{M}(t)  \cdot \vec{M}(0) \rangle$ is small as compared to $\left \langle \left| \vec{M} \right| ^2 \right \rangle$. Therefore, we deduce that MSD$_M$ saturates as 
\begin{equation} \label{eq:saturate}
\begin{split}
\text{MSD}_M(t) & = \left \langle \left| \vec{M}(t)\right|^2 + \left| \vec{M}(0) \right|^2 - 2 \vec{M}(t) \cdot \vec{M}(0) \right \rangle \\
 & \approx 2\langle M^2 \rangle \approx 2 k_b T L^2 \chi.
\end{split}
\end{equation}

\begin{figure}
\includegraphics[width=8.0cm]{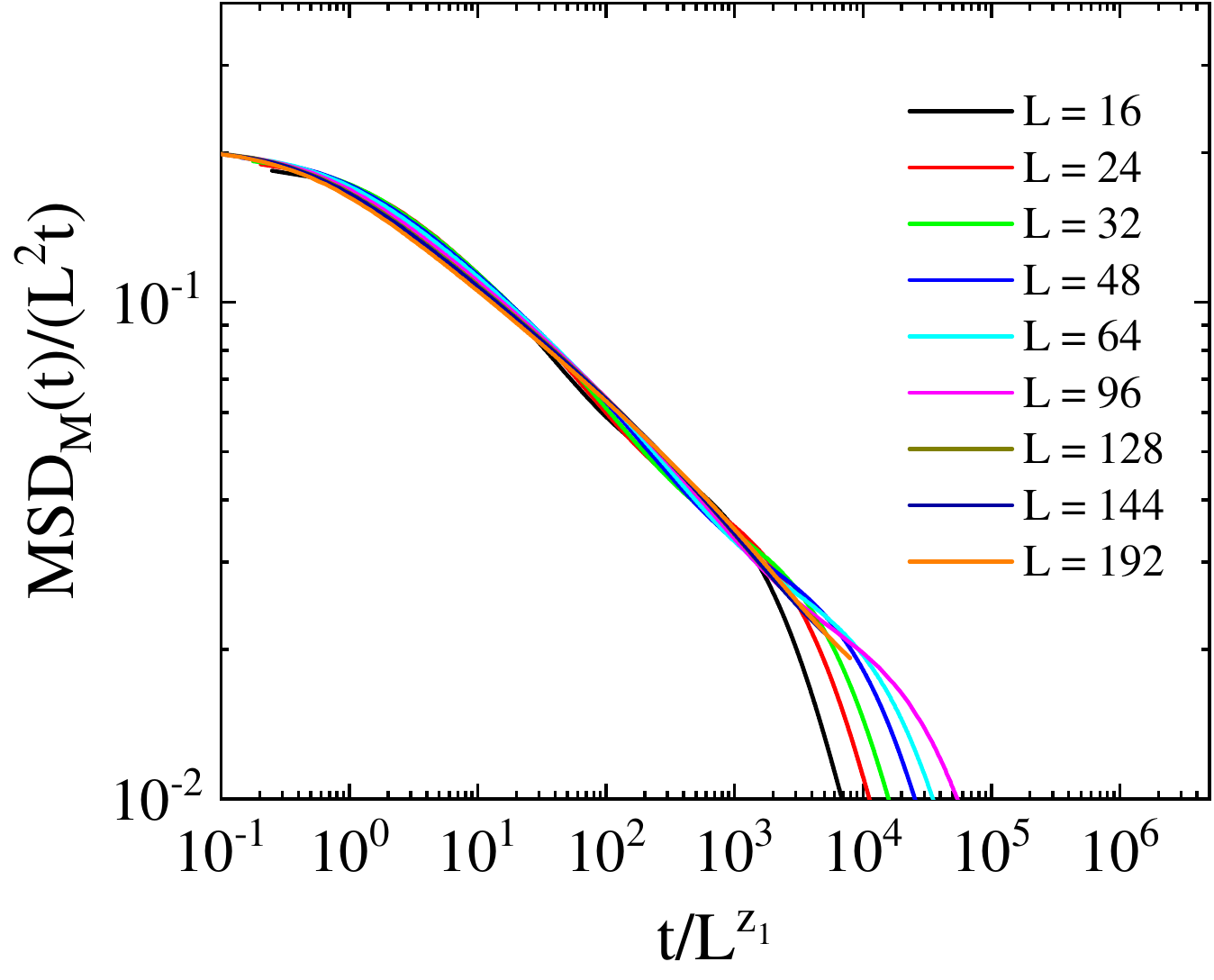}
\caption{\label{Fig2} Data collapse of the MSD$_M(t)$ curves for the full range of system sizes studied around the first crossover $L^{z_1}$ with a scaling form of MSD$_M(t)/(L^2 t) \sim t/L^{z_1}$, where $z_1$ is $0.50(5)$. The MSD$_M(t)$ shifts from normal ($\sim L^2t$) to anomalous diffusion ($\sim L^{2+z_1-\alpha z_1}t^\alpha$) at $t = L^{z_1}$.}
\end{figure}

Rather than an operational procedure, the dynamics can also be articulated from the application of the transition matrix $\mathcal{A}$ to the state vector $\vec{S}$. This is clearly a dysfunctional formulation due to the fact that $\mathcal{A}$ is a sparse $2^{L^2} \times 2^{L^2}$ matrix, but nevertheless useful for the sake of argument. This transition matrix has an eigenvalue of $e_{0} = 1$, with an eigenvector in which each element lies the likelihood of that state (the Boltzmann distribution). It also has a second-highest eigenvalue $e_{1} \approx 1$, which determines the final exponential decay of the autocorrelation function. At long times $t$, the dynamical matrix is applied $t L^2$ times. Thus, expressed in $A$ the dynamics can be written as $C_M(t) =\langle \vec{S}_t \mathcal{A}^{tL^2} \vec{S}_0 \rangle$. For long times, the decay of the autocorrelation function is dominated by the largest non-zero eigenvector and eigenvalue $C_M(t) \sim e_1^{tL^2} \sim \exp{[-t/\tau_2]}$, where $\tau_2 = -L^2 \ln{(e_1)}$. Clearly, it is almost impossible to numerically retrieve $\tau_2$ via $e_1$, unless $L$ is a very small number. However, the whole construction provides a solid argument suggesting that the magnetization autocorrelation function will eventually decay exponentially at long times for finite $L$. 

At the same time, the intermediate regime plays also a crucial role in connecting the short- and long-time regimes monotonically. The numerical data suggest that this happens via anomalous diffusion, i.e., MSD$_M(t) \sim t^\alpha$, whereas the autocorrelation function seems to decay as a stretched-exponential with the same anomalous exponent $\alpha$.

\begin{figure}
\includegraphics[width=8.0cm]{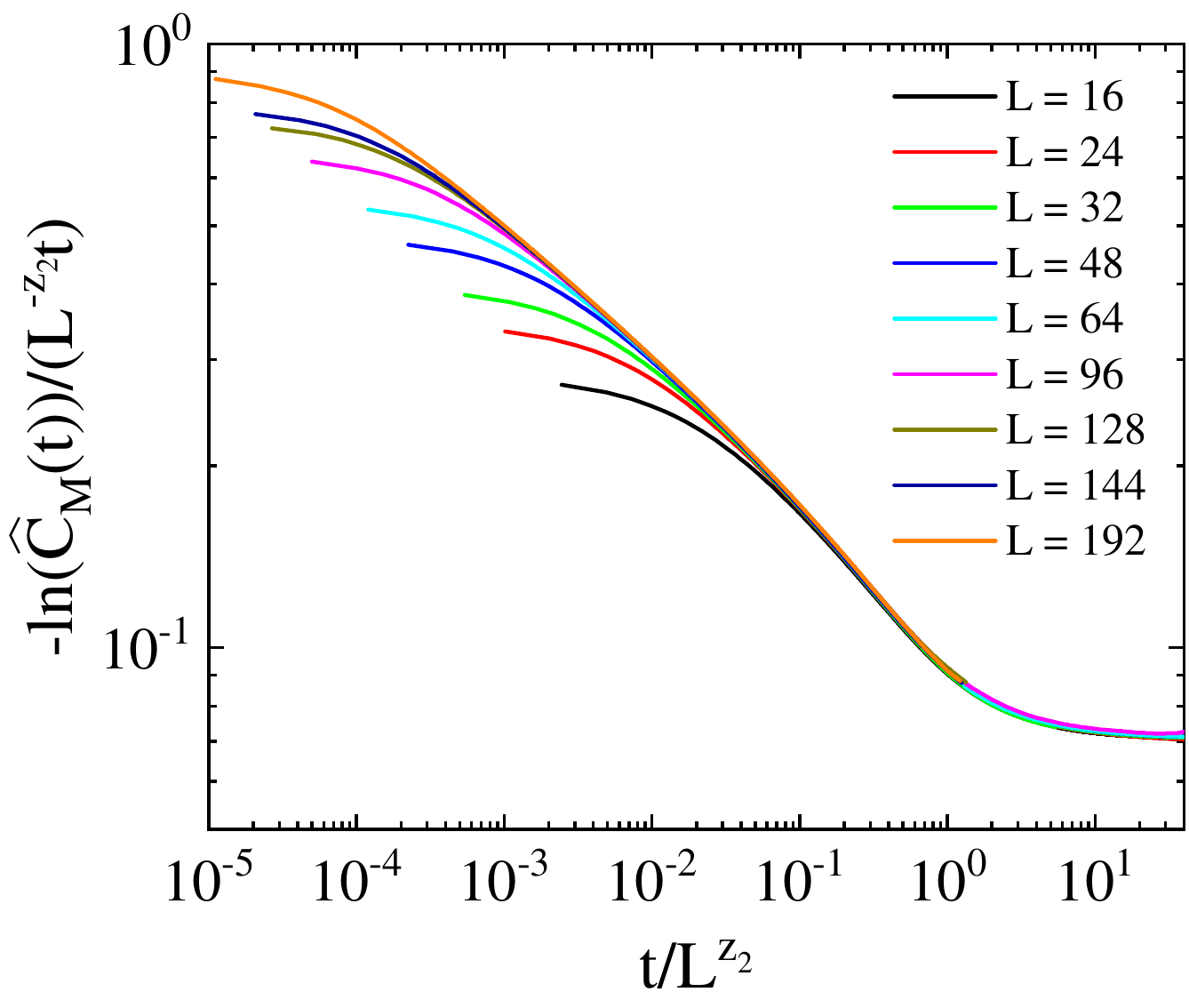}
\caption{\label{Fig3} Data collapse of $-\ln{(\hat{C}_M(t))}$ for the full range of system sizes studied around the second crossover with a scaling form $-\ln(\hat{C}_M(t))/L^{-z_2}t \sim t/L^{z_2}$, where $z_2 = 2.17(1)$.}
\end{figure}

As in Ref.~\cite{liu23}, the main target of this work is to obtain, via finite-size scaling approaches~\cite{amit_book}, access to the dynamic exponents $z_1$ and $z_2$ that mark the crossover behavior between the three aforementioned regimes, as well as the anomalous exponent $\alpha$. Figure~\ref{Fig2} imprints the collapse of MSD$_M(t)$ curves for all system sizes studied in the area around the first transition point, obtained for $z_1 = 0.50(5)$, in agreement with the result $z_1 = 1/2$ for the Ising ferromagnet~\cite{liu23}. At the intermediate regime of this plot the curve is expected to decay as $\sim t^{\alpha-1}$, leading to the numerical estimate of $\alpha = 0.74(4)$. Figure~\ref{Fig3} now presents an analogue to Fig.~\ref{Fig2}, this time a collapse of the curves around the second transition point. This is established by plotting $-\ln{(\hat{C}_M(t))}/(L^{-z_2}t)$ as a function of $t/L^{z_2}$, where in this case $z_{2} = 2.17(1)$, in agreement with the estimate $2.17(4)$ by Tang and Landau~\cite{tang87}, but even more importantly with the result $z_2 = 13/6 \approx 2.1667$~\cite{nightingale96,liu23} for the Ising ferromagnet.

The intermediate regime for MSD$_M$ initiates at time $\tau_1 \sim L^{z_1}$ at a value of $\langle (\Delta M)^2 \rangle \sim L^{2+z_1}$, followed by a power-law increase controlled by the exponent $\alpha$, and finally reaching its saturation value $\sim L^{2+\gamma/\nu}$ at time $\tau_2 \sim L^{z_2}$. Following the assumptions of Ref.~\cite{liu23} for a single power-law dependence in the intermediate regime, the anomalous exponent $\alpha$ can be expressed via
\begin{equation}
\label{eq:exporel}
\alpha = (\gamma/\nu-z_{1})/(z_{2} - z_{1}).
\end{equation}
Inserting the values $\gamma/\nu = 26/15$, $z_1 = 1/2$, and $z_2 = 13/6$ in the above Eq.~\eqref{eq:exporel} we obtain the analytical result $\alpha = 37/50 = 0.74$, in excellent agreement with the numerical estimate $0.74(4)$ obtained from the scaling analysis of Fig.~\ref{Fig2} (note the value $\alpha=3/4$ for the Ising case~\cite{liu23}). This is another strong evidence in favor of the dynamical universality hypothesis between Ising and Potts models, as exemplified below in more detail. 

\subsection{A direct comparison of Ising and Potts models}
\label{sec:comparison}

Since the values obtained above for the dynamic exponents of the three-state Potts model are up to a good numerical accuracy compatible to those of the Ising ferromagnet~\cite{liu23}, we perform here an additional direct comparison of the numerical data for both models, considering equally the mean-squared deviation of the magnetization and the magnetization autocorrelation functions. 

\begin{figure}
\includegraphics[width=6.9cm]{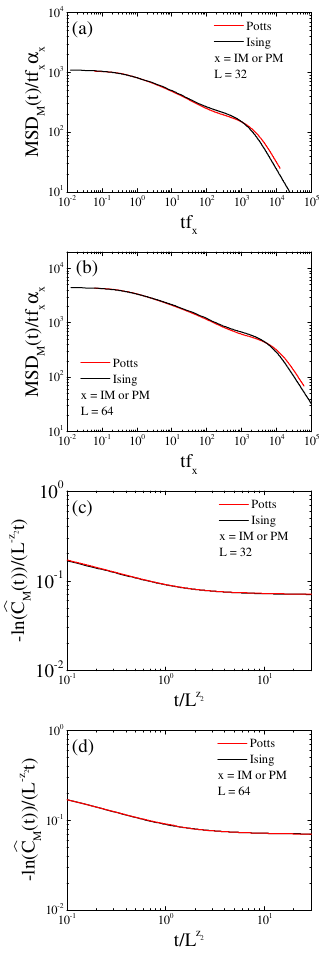}
\caption{\label{Fig4} Comparative results for the 
two-dimensional three-state Potts (red curves) and Ising (black curves) models. Panels (a) and (b) showcase the data collapse of the MSD$_M$(t) curves with a scaling form of MSD$_M(t)/(t f_x \alpha_x) \sim tf_x$, where $f_x$ and $\alpha_x$ are some arbitrary constants for Ising and Potts models, respectively. Panels (c) and (d) present the collapse of $-\ln{(\hat{C}_M(t))}$ around the second crossover with a scaling form $-\ln(\hat{C}_M(t))/L^{-z_2}t \sim t/L^{z_2}$. Numerical data for two system sizes are shown, namely $L = 32$ (upper panels) and $L = 64$ (lower panels).}
\end{figure}

In Fig.~\ref{Fig4}(a)--(b) we focus on the relative values of the first exponent $z_1$. For a more meaningful comparison, we exploit the knowledge that, at short times, the MSD$_M(t)$ for the Potts model is smaller than that for the Ising model simply because the acceptance ratio for the proposed moves is larger -- numerically we find this ratio to be $f_{\rm P} = 0.33$ (Potts) and $f_{I} = 0.14$ (Ising). Furthermore, the change in the squared magnetization due to a single spin-flip corresponds to $\frac{3}{2}$ and $4$ for the Potts and Ising models, respectively. After removing both of these relatively trivial effects, the curves of Fig.~\ref{Fig4}(a)--(b) show a remarkable similarity, to the degree that numerically we cannot determine which of the two models has a larger exponent $z_1$; our numerical estimation is that if there is a difference in $z_1$, it does not exceed the order of $\sim 10^{-3}$. Subsequently we elaborate on the relative values of the second exponent $z_2$ in Fig.~\ref{Fig4}(c)--(d). Here, we apply arbitrary scaling factors along the horizontal and vertical axes and plot $-\ln(\hat{C}_M(t))/L^{-z_2}t$ vs. $t/L^{z_2}$. Again, the curves show an excellent matching, indicating that the difference in the numerical value of the dynamic exponent $z_2$ between the two models is negligible, and thus undetectable in our simulations. 

Finally, as mentioned in Sec.~\ref{sec:intro}, the exponent ratio $\gamma/\nu$ which is known exactly for both models in equilibrium, appears to be only slightly different; note the values $7/4 = 1.75$ (Ising ferromagnet) and $26/15 \approx 1.733$ (three-state Potts model). When combining Eq.~\eqref{eq:exporel} with the numerically indistinguishable values for the dynamic exponents $z_1$ and $z_2$ obtained in the current work and in Ref.~\cite{liu23} one would expect a slight difference in the anomalous exponent $\alpha$. This is indeed the case, as is clearly visible from Fig.~\ref{Fig4}(a)--(b).

\begin{figure}
\includegraphics[width=8.35cm]{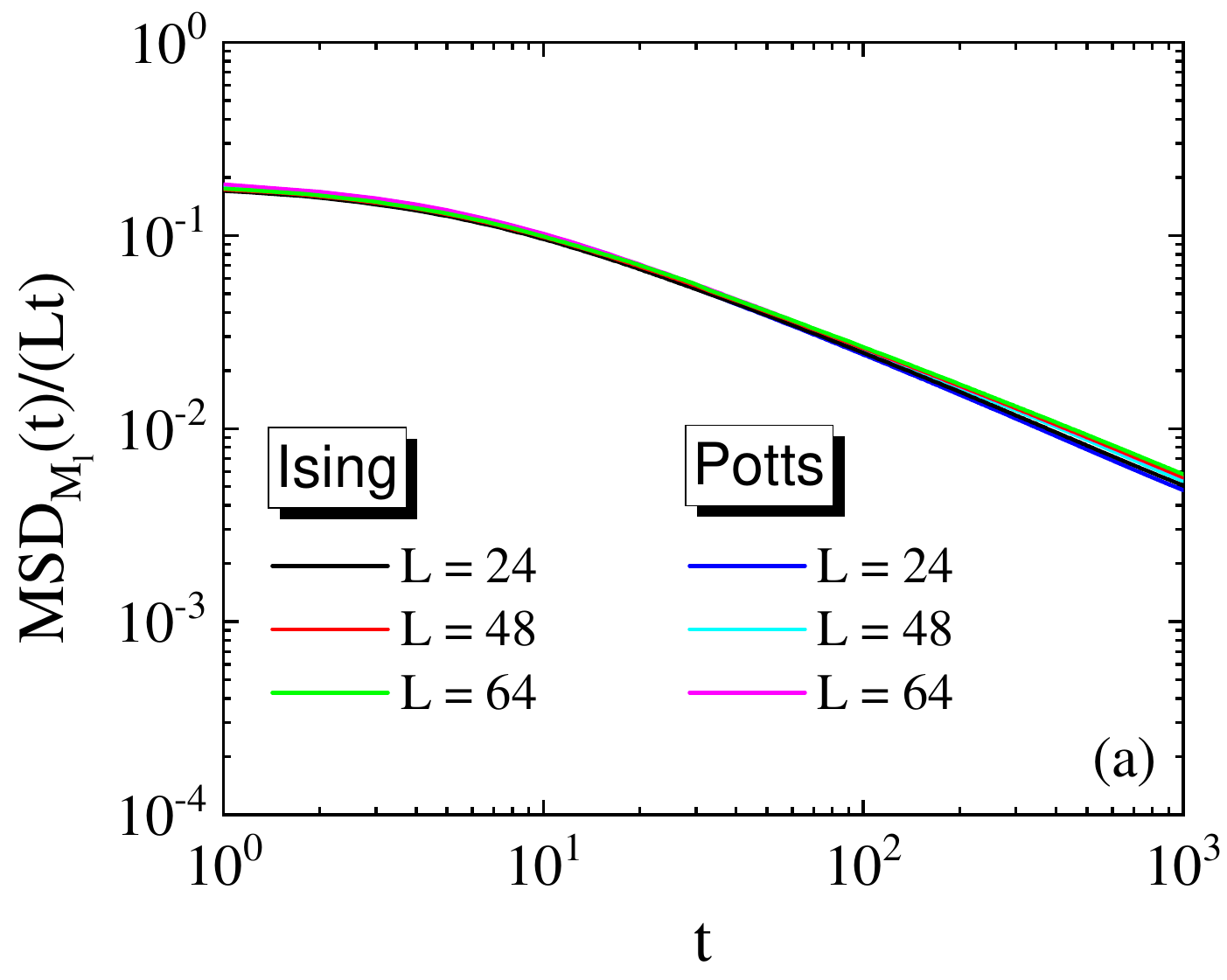}
\includegraphics[width=8.0cm]{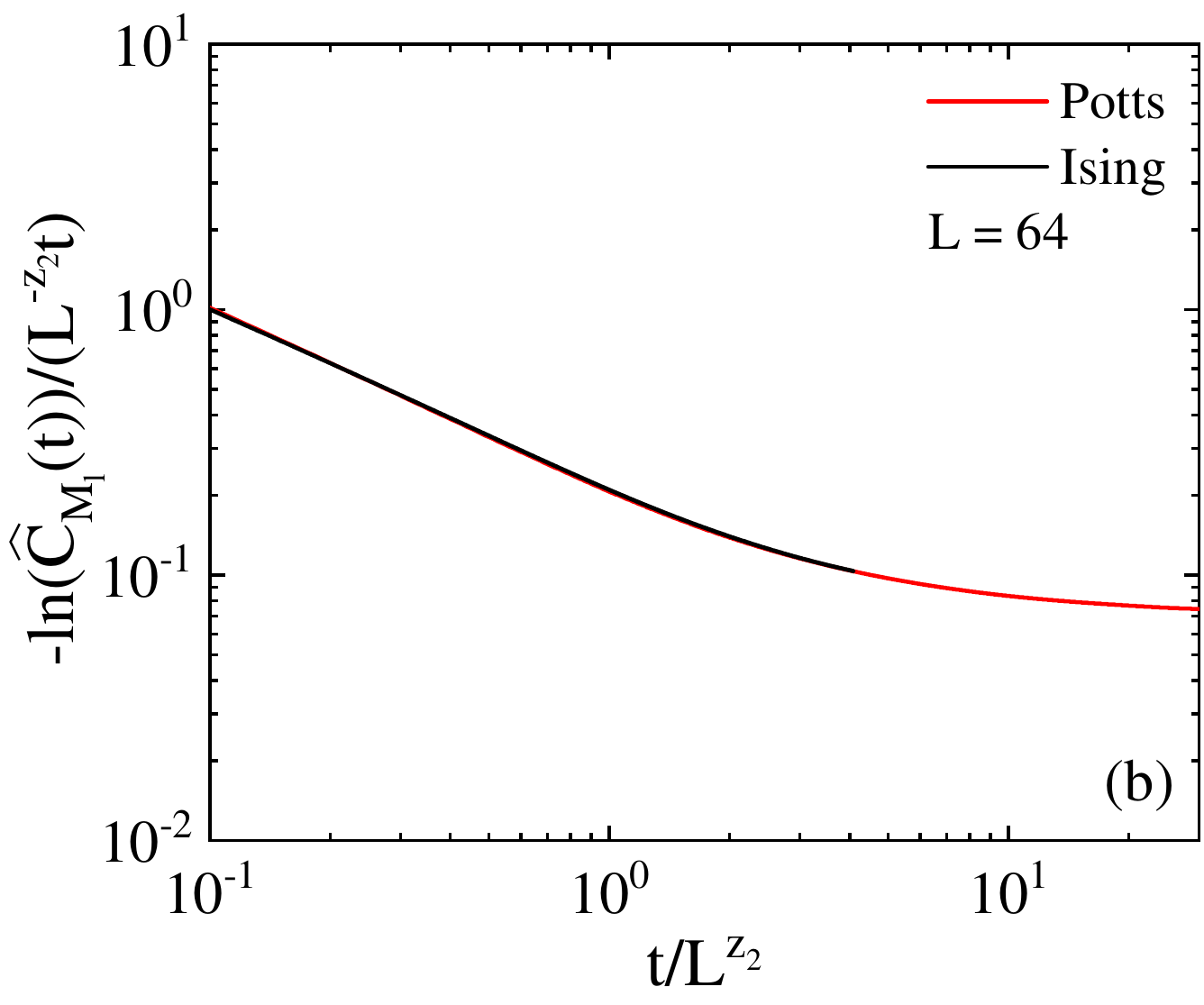}
\caption{\label{Fig5} Results based on the analysis of the line magnetization for both the three-state Potts and Ising models. (a) Data collapse of the MSD$_{M_{l}}$(t) curves with a scaling form of MSD$_{M_{l}}(t)/(Lt) \sim t/L^{z_{1}^{(l)}}$ around the first crossover regime for $L = 24$, $48$, and $64$. (b) Data collapse of $-\ln{(\hat{C}_{M_{l}}(t))}$ according to the scaling form $-\ln(\hat{C}_{M_{l}}(t))/(L^{-z_{2}^{(l)}}t) \sim t/L^{z_{2}^{(l)}}$ for $L = 64$ around the second crossover regime.}
\end{figure}

\subsection{Line magnetization}
\label{sec:line}

Besides the most studied bulk magnetization, another convenient observable is the line magnetization, which can also provide insightful results pertaining to the universality aspects of criticality. To define the line magnetization we need to switch notation from the generic coordinate $i$ used above, to the location $(x,y)$, so that the new observable is simply the sum of spins along the sites with the same $x$-coordinate (or alternatively, $y$-coordinate). For convenience, we now focus on the line magnetization for the collection of the $x = 0$ sites, given by
\begin{equation}
\vec{M}_l(x=0) = \sum_{y=0}^{L-1} \vec{M}(0,y).    
\end{equation}
In full analogy to the bulk magnetization, both the mean-squared deviation (MSD$_{M_{l}}(t)$) and the normalized autocorrelation function ($\hat{C}_{M_{l}}(t)$) of the line magnetization $M_{l}$ can be defined. In fact, our simulations revealed the same dynamical behavior featuring three distinct regimes separated by two crossover behaviors distinguished at times $t\sim L^{z_1^{(l)}}$ and $t\sim L^{z_2^{(l)}}$: (i) A first short-time regime where the MSD$_{M_{l}}$ increases linearly with time and the autocorrelation function remains constant, (ii) a second intermediate regime where the MSD$_{M_{l}}$ increases as a power law, but the autocorrelation function decreases as a stretched exponential, and (iii) a final asymptotic regime where the MSD$_{M_{l}}$ saturates with the autocorrelation function decaying exponentially. 

As in Sec.~\ref{sec:bulk} above, we determined numerically the dynamic exponents $z_1^{(l)}$ and $z_2^{(l)}$, as well as the anomalous exponent $\alpha^{(l)}$ via data collapses of the available numerical data. In this case, we considered systems with linear sizes $L\in \{24,48,64\}$. The collapsed curves are shown in Fig.~\ref{Fig5} for both the three-state Potts and Ising models. The resulting values are listed below: (i) $z_1^{(l)} = 0$ for Ising and Potts, clearly different from the values obtained for the bulk magnetization, (ii) $z_2^{(l)} = 2.17(1)$, numerically indistinguishable from the estimates obtained for the bulk magnetization for both models, and (iii) $\alpha^{(l)} =0.35(1)$ and $0.34(1)$ for the Potts and Ising models, respectively. This slight difference is visible in Fig.~\ref{Fig5}(a) from the minor variation in the relevant slopes. Note that the Ising result $\alpha^{(l)} = 0.34(1)$ is in agreement with the earlier work of Ref.~\cite{zhong18}, and that for the line magnetization, continuity requires $\alpha^{(l)}=(\gamma/\nu-1-z_1^{(l)})/(z_2^{(l)}-z_1^{(l)})$, as the saturation of the MSD$_{M_{l}}$ folows the scaling of the form $\sim L^{\gamma/\nu+1}$.

\section{Concluding Remarks}
\label{sec:summary}

We analyzed the results of extensive simulations of the two-dimensional three-state Potts model with local spin-flip dynamics. We scrutinized the time evolution of the mean-squared deviation and autocorrelation function of the bulk and line magnetizations, featuring three dynamical regimes separated by two crossover times at $\tau_1 \sim L^{z_1}$ and $\tau_2 \sim L^{z_2}$. In the short-time regime, the mean-squared deviation shows ordinary diffusive behavior and the autocorrelation function exponential decay. In the second intermediate regime the mean-squared deviation is characterized by anomalous diffusive behavior and the autocorrelation function decays in a stretched-exponential way. Finally, in the third late-time regime the mean-squared deviation saturates at a constant value while the autocorrelation function again decays exponentially. This intricate behavior was originally highlighted in Ref.~\cite{liu23} for the square-lattice Ising ferromagnet and its bulk magnetization.

In particular, the second crossover to the exponential decay of the autocorrelation function is well documented in the literature for the bulk magnetization of the two-dimensional Ising model. Nightingale and Bl\"ote elucidated that this decay sets in at a time determined by the dynamic critical exponent $z = 2.1665(12)$~\cite{nightingale96}. Our results are in agreement with this value, but extend also to the line magnetization of the Ising model, and more surprisingly to the bulk and line magnetizations of the three-state Potts model.
On the other hand, the first crossover has only recently been disclosed by our group~\cite{liu23} for the bulk magnetization of the square-lattice Ising ferromagnet, and was found to occur at a time determined by a new dynamic exponent $z_1\approx 0.5$. In the current work, we find that this exponent shares the same value for the bulk magnetization of the three-state Potts model, similar to the exponent $z_2$. However, the same analysis based on a different observable, namely the line magnetization, suggested the value $z_1^{(l)} \approx 0$ for both Ising and Potts models. At the moment, we don't have any theoretical argument for this numerically observed behavior.

In the intermediate regime, the mean-squared deviation of the magnetization shows power-law behavior with an anomalous exponent $\alpha$, and related to this, the autocorrelation function a stretched-exponential decay with the same exponent. Continuity sets a relation between $\gamma/\nu$ (the ratio of two equilibrium critical exponents), $z_1$, $z_2$ and $\alpha$; see Eq.~\eqref{eq:exporel}. Since $z_1$ and $z_2$ are shared between the Ising and three-state Potts models but $\gamma/\nu$ is not, it is not surprising that $\alpha$ takes different values between the two models. This result pertains to both the bulk and line magnetizations considered.

To conclude, we provided numerical evidence suggesting that at two dimensions the three-state Potts model and the Ising ferromagnet, which belong to distinct equilibrium universality classes, share within our numerical accuracy their dynamical critical exponents $z_1$ and $z_2$. This result has been obtained using heat-bath dynamics but we expect it to hold also for other types of single spin-flip dynamics as well, such as Metropolis, Glauber, and others. At this stage, it would be very interesting to place the results of the present work in a more general framework of universality but this would require an extensive testing of the current protocol against different spin models, and perhaps also different algorithm dynamics (including the most commonly used cluster algorithms). Work in this direction is currently under way. We hope that our work will stimulate additional research on the field of dynamical critical phenomena providing a more rigorous theoretical ground to accommodate for the reported numerical results.

\begin{acknowledgments}
We would like to thank Peter Grassberger and Martin Hasenbusch for fruitful correspondence on the problem of dynamical critical phenomena. We acknowledge the provision of computing time  from T\"{U}B\.{I}TAK ULAKB\.{I}M (Turkish agency), High Performance and Grid Computing Center (TRUBA Resources). The work of N.G. Fytas was supported by the  Engineering and Physical Sciences Research Council (grant EP/X026116/1 is acknowledged).
\end{acknowledgments}


\begin{thebibliography}{199}

\bibitem{fisher74} M.E. Fisher, Rev. Mod. Phys. {\bf 46}, 597 (1974).

\bibitem{fisher98} M.E. Fisher, Rev. Mod. Phys. {\bf 70}, 653 (1998).

\bibitem{wilson71} K.G. Wilson, Phys. Rev. B {\bf 4}, 3174 (1971).

\bibitem{landau_book} D. P. Landau and K. Binder, \textit{A Guide to Monte Carlo Simulations in Statistical Physics} (Cambridge University Press, Cambridge, England, 2000).

\bibitem{barkema_book} M.E.J. Newman and G.T. Barkema, \textit{Monte Carlo methods in Statistical Physics} (Clarendon Press, 1999).

\bibitem{amit_book} D.J. Amit and V. Mart\'in-Mayor, \textit{Field Theory, the Renormalization Group and Critical Phenomena}, 3rd ed. (World Scientific, Singapore, 2005).

\bibitem{Ising25} E. Ising, Z. Physik {\bf 31}, 253 (1925).

\bibitem{potts} R.B. Potts, Mathematical Proceedings of the Cambridge Philosophical Society, {\bf 48}, 106 (1952) 

\bibitem{wu82} F.Y. Wu, Rev. Mod. Phys. {\bf 54}, 235 (1982).

\bibitem{hohenberg77} P.C. Hohenberg and B.I. Halperin, Rev. Mod. Phys. {\bf 49}, 435 (1977).

\bibitem{folk06} R. Folk and G. Moser, J. Phys. A: Math. Gen. {\bf 39}, R208 (2006).

\bibitem{hasenbusch07} M. Hasenbusch, A. Pelissetto, and E. Vicari, J. Stat. Mech. (2007) P11009.

\bibitem{zhong20} W. Zhong, G.T. Barkema, and D. Panja, Phys. Rev. E {\bf 102}, 022132 (2020).

\bibitem{nightingale96} M.P. Nightingale and H.W.J. Bl{\"o}te, Phys. Rev. Lett. {\bf 76}, 4548 (1996).

\bibitem{hasenbusch20} M. Hasenbusch, Phys. Rev. E {\bf 101}, 022126 (2020).

\bibitem{janus} M. Baity-Jesi, \textit{et al.} (Janus Collaboration), Phys. Rev. Lett. {\bf 120}, 267203 (2018).

\bibitem{liu23} Z. Liu, E. Vatansever, G.T. Barkema, and N.G. Fytas, Phys. Rev. E {\bf 108}, 034118 (2023).

\bibitem{janssen89} H.K. Janssen, B. Schaub, and B. Schmittmann, Z. Physik B - Condensed Matter {\bf 73}, 539(1989).

\bibitem{tobochnik81} J. Tobochnik and C. Jayaprakash, Phys. Rev. B {\bf 25}, 4893 (1981).

\bibitem{arcangelis86} L de Arcangelis and N. Jan, J. Phys. A: Math. Gen. {\bf 19}, L1179 (1986).

\bibitem{tang87} S. Tang and D.P. Landau, Phys. Rev. B {\bf 36}, 567 (1987).
 
\bibitem{schulke95} L. Sch{\"u}lke and B. Zheng, Phys. Lett. A {\bf 204}, 295 (1995).

\bibitem{schulke96} L. Sch{\"u}lke and B. Zheng, Phys. Lett. A {\bf 214}, 81 (1996).

\bibitem{murase08} Y. Murase and N. Ito, J. Phys. Soc. Jpn. {\bf 77}, 014002 (2008). 

\bibitem{martinelli99} F. Martinelli, \textit{Lectures on Glauber Dynamics for Discrete Spin Models} In: Bernard, P. (eds) Lectures on Probability Theory and Statistics. Lecture Notes in Mathematics, vol 1717. Springer, Berlin, Heidelberg).

\bibitem{randall00} D. Randall and P. Tetali, J. Math. Phys. {41}, 1598 (2000).

\bibitem{coulon04} C. Coulon, R. Cl{\'e}rac, L. Lecren, W. Wernsdorfer, and H. Miyasaka, Phys. Rev. B {\bf 69}, 132408 (2004).

\bibitem{grandi96} B.C.S. Grandi and W. Figueiredo, 
Phys. Rev. E {\bf 53}, 5484 (1996).

\bibitem{smedt03} G. De Smedt and C. Godreche, Eur. Phys. J. B {\bf 32}, 215 (2003).

\bibitem{godreche04} C. Godreche, F. Krzaka{\l}a, Florent, and F. Ricci-Tersenghi, J. Stat. Mech.: Theory and Exp. (2004) P04007.

\bibitem{coddington92} P.D. Coddington, D. Paul, and C.F. Baillie, Phys. Rev. Lett. {\bf 68}, 962 (1992).

\bibitem{rieger99} H. Rieger and N. Kawashima, Eur. Phys. J. B  {\bf 9}, 233 (1999).

\bibitem{bloete02} H.W.J. Bl{\"o}te and Y. Deng, Phys. Rev. E {\bf 66}, 066110 (2002).

\bibitem{zhong18} W. Zhong, D. Panja, G.T. Barkema, and R.C. Ball, Phys. Rev. E {\bf 98}, 012124 (2018).

\end{thebibliography}
\end{document}